  \documentclass[aps,prb,preprint,groupedaddress]{revtex4-2}
\usepackage{bm}
\usepackage{amsmath}
\usepackage{amssymb}
\usepackage{graphicx}
\usepackage{xcolor}
\begin{document}
%--
% Use the \preprint command to place your local institutional report
% number in the upper righthand corner of the title page in preprint mode.
% Multiple \preprint commands are allowed.
% Use the 'preprintnumbers' class option to override journal defaults
% to display numbers if necessary
%\preprint{}
%--
%--
% My definitions / start
%--
\def\mysecskip{\medskip}
\def\opf{\Omega}
\def\opfe{{\overline{\opf}}}
\def\khat{\hat{k}}
\def\rhat{\hat{r}}
\def\zhat{\hat{0}}
\def\kvec{{\mathbf k}}
\def\qvec{{\mathbf q}}
\def\qscaleV{\bm{\kappa}}
\def\qscaleS{{\kappa}}
\def\zedvec{{\mathbf z}}
\def\rvec{{\mathbf r}}
\def\Rvec{{\mathbf R}}
\def\udisvec{{\mathbf u}}
\def\vdisvec{{\mathbf v}}
\def\udisscaF{U}
\def\udisvecF{{\mathbf U}}
\def\zvec{{\mathbf 0}}
\def\Dim{D}
\def\dim{d}
\def\efe{{\cal H}}
\def\pno{N}
\def\vol{V}
\def\ctp{\tau}
\def\hrs{\rm HRS}
\def\lrs{\rm LRS}
\def\cocon{g}
\def\locfrac{Q}
\def\dist{{\cal P}}
\def\scadist{\Pi}
\def\scat{\theta}
\def\sds{{\cal S}}
\def\hatpi{\hat{\Pi}}
\def\hatth{\hat{\scat}}
\def\tcomb{\mathbb T}
\def\barexi{\xi_{0}}
\def\crlide{^{\small\chi}\!N}
%--
\def\ie{{\it i.e.\/}}
\def\eg{{\it e.g.\/}}
\def\viz{{\it viz.\/}}
\def\via{{\it via\/}}
%--
\def\eas{{equilibrium amorphous solid state}}
%--
\def\myQsum{\bm{\lambda}}
\def\myQQsum{\bm{\Lambda}}
%--
\def\MUvec{\bm{\mu}}
\def\rmsFl{{\cal F}}
\def\rmsCo{{\cal C}}
\def\MUvecDUM{\bm{\mu}}
\def\rmsFlDUM{{\cal F}}
\def\rmsCoDUM{{\cal C}}
\def\ProbCharsLarge{{\cal P}}
\def\mynewhalf{\tfrac{1}{2}}
%--
\def\MUvec{\bm{\mu}}
\def\rmsFl{{\cal F}}
\def\rmsCo{{\cal C}}
%--
  \def\MUvecDUM{{\bf m}}
  \def\rmsFlDUM{{\mathbb{F}}}
  \def\rmsCoDUM{{\mathbb{C}}}
%--
\def\ProbCharsLarge{{\mathbb{P}}}
\def\distPHY{{\widetilde{\ProbCharsLarge}}}
\def\Ucorten{{\mathbb{G}}}
\def\mynewhalf{\tfrac{1}{2}}
%--
\def\loc{L}
\def\locfrac{Q}
  \def\AVtherm{{}}
  \def\AVdisor{{}}
%--
\newcommand{\overbar}[1]{\mkern 1.5mu\overline{\mkern-1.5mu#1\mkern-1.5mu}\mkern 1.5mu}
\makeatletter
\newcommand{\Vast}{\bBigg@{3}}
\makeatother
\newcommand*{\Scale}[2][4]{\scalebox{#1}{$#2$}}
\def\mybar{\overbar}
% \def\mybar{\overline}
%--
\def\cartD{d}
\def\cartDbar{\mybar{d}}
%--
% \def\mynewcolor{\textcolor{blue}}
\def\mynewcolor{\textcolor{black}}
%   remember to open and close
%       the text to be colored using
%       {...} %endmynewcolor
%--
% \underset{\kappa\to 0}{=} [puts one below the other; stacks]
%--
% My definitions / end
%--
%--
%Title of paper
\title{Statistical field theory of equilibrium amorphous solids
and the intrinsic heterogeneity distributions that characterize them}
%--
% \title{Vulcanized matter: An exemplary equilibrium amorphous solid}
% \title{Vulcanized matter: The exemplary equilibrium amorphous solid}
% \title{The equilibrium amorphous solid state}
%------------------------------------------
% * * * * * * * * * * * * * * * * * * * * *
% * * * * * * * * * * * * * * * * * * * * *
% * * * * * * * * Source notes  * * * * * *
% * * * * * * * * * * * * * * * * * * * * *
%    For the scale-dependent elasticity computation
%    NB-Vulcan/NB-Vulcan-Goldstone-branch
%             /GM-energy/2nd order u exp
%   For the free energy and order parameter on which it is based
%-------------------------------------------

% repeat the \author .. \affiliation  etc. as needed
% \email, \thanks, \homepage, \altaffiliation all apply to the current
% author. Explanatory text should go in the []'s, actual e-mail
% address or url should go in the {}'s for \email and \homepage.
% Please use the appropriate macro foreach each type of information

% \affiliation command applies to all authors since the last
% \affiliation command. The \affiliation command should follow the
% other information
% \affiliation can be followed by \email, \homepage, \thanks as well.

\author{Paul M.~Goldbart}
\email[]{paul.goldbart@stonybrook.edu}
%\homepage[]{Your web page}
%\thanks{}
%\altaffiliation{}
\affiliation{Department of Physics and Astronomy, Stony Brook University,
Stony Brook, NY~11794, USA}

%Collaboration name if desired (requires use of superscriptaddress
%option in \documentclass). \noaffiliation is required (may also be
%used with the \author command).
%\collaboration can be followed by \email, \homepage, \thanks as well.
%\collaboration{}
%\noaffiliation

%    \date{April 1, 2025}
% ** \date{\today}

\begin{abstract}
%--
A rich variety of amorphous solids are found throughout nature and technology, including those formed \via\ the vulcanization of long, flexible molecules.
%--
A special class of these solids -- the ones that feature a wide separation between the long timescales over which constraints in them release and the much shorter timescales over which their unconstrained degrees of freedom relax -- exhibit states of thermodynamic equilibrium, and are therefore amenable to the framework of equilibrium statistical physics.
%--
The statistical physics of equilibrium amorphous solid-forming systems may be approached at several levels of detail and, accordingly, the conclusions of these approaches have distinct levels of specificity.
%--
The approach reviewed here is the least detailed and thus the most general: statistical field theory.
%--
An overview is given of the core ingredients and key results of the statistical field theory of equilibrium amorphous solids.
%--
The field at the center of this theory is motivated by the aim of detecting and diagnosing the amorphous solid state.
%--
Its form turns out to be rather unusual, in ways that are essential for its application.
%--
It is therefore examined in detail, as is the form of the statistical field theory that controls the field's average value and fluctuations.
%--
What this theory predicts -- and can predict -- for the equilibrium properties of amorphous solids is then discussed, including:
%--
the transition from the liquid to the amorphous solid state and the structure and heterogeneity of the resulting solid;
the impact of fluctuations on the transition and connections with percolation theory;
the pattern of translational symmetry-breaking and the nature of the elasticity that results from it, including its dependence on lengthscale and what this reveals about the solid's heterogeneity; and
field-fluctuation correlations and the information they provide.
%--
Along the way, emphasis will be placed on the idea, particular to amorphous solids, that their equilibrium states are most naturally characterized in terms of statistical distributions that describe the intrinsic spatial heterogeneity of the thermal motions of their constituents.
%--
The field on which the statistical field theory is based has an internal structure to it that encodes this information in a subtle manner, \via\ the wave-vector dependencies of the average field and its fluctuation correlations.
%--
This review concludes with some reflections on the applicability -- or otherwise -- of the ideas and results it explores to a variety of amorphous solids and related systems.
\end{abstract}%

% insert suggested keywords - APS authors don't need to do this
%\keywords{}

% \pacs{03.65.Pm, 05.30.Fk, 36.20.Fz, 61.30.Vx}
% \pacs{**}

%\maketitle must follow title, authors, abstract, and keywords
\maketitle

% body of paper here - Use proper section commands
% References should be done using the \cite, \ref, and \label commands
%
%--
%                \hfil\break
%--
%--
% ** Key pointers from ARCMS...
% A critical appraisal of the signiﬁcant, rather than the total,
% literature in the ﬁeld.
% May include your own work (even if unpublished).
% Should be useful to specialists as well as teachers and scholars
% from other areas.
% Should emphasize where research in a given area should go, as
% well as where it has been, such that it will inﬂuence the
% future course of knowledge.
%--
%--
% ** What figures should we have (look at past talks)
% Macromolecules / Crosslinks / Deam-Edwards distribution /
% Localized particles / one- and two-particle properties /
% Semi-microscopic replica field theory /
% How the symmetry breaks / Goldstone excitations
% Stiffness / Scale-dpendent stiffness
%--
%--
\newpage
%--
\mysecskip
% \section{Scope\label{sec:intro}}
\noindent{\bf What is vulcanized matter and what are equilibrium amorphous solids?\/}:
My aim with this article is to give an entr{\'e}e into the physics of soft solids of the type that most commonly arise when fluid systems consisting of many long, flexible, fully mobile molecules are subjected to vulcanization.
%--
This is the process in which essentially permanent chemical bonds, often called crosslinks, are introduced between some random fraction of the molecules that happen to be in near-contact with one another at the time of vulcanization.
%--
One can think of the constraints resulting from the vulcanization process as locking in elements of the molecular-level structural organization that were present in the liquid while vulcanization was taking place.
%--
However, whilst aspects of liquid-like structure are indeed maintained in a sense to be made precise below, the post-vulcanization system is not necessarily liquid.
%--
Rather, the introduction of a sufficiently large number of of random bonds -- only of order one per molecule, but enough to establish a macroscopic network of molecules -- converts the system from a liquid into a solid.

Viewed microscopically, the emergent characteristic of this solid is that allowed relaxations are inadequate to foster liquidity.
%--
At least a fraction of the molecules no longer wander throughout the container (as they do in a liquid, given sufficient time).
%--
Instead, they undergo finite-ranged thermal fluctuations about the mean positions to which they are bound.
%--
These mean positions are themselves arranged randomly in space, reflecting the disorganized structure of the fluid from which the solid has descended, so that the collection of mean positions across the system appears amorphous (\ie, has no evident form, in contrast with the case of a crystal~\cite{ref:crystal}).
%--
The spatial extents of the molecules\rq\ position-fluctuations are random, too, as a result of the random environments that the bound molecules inhabit.
%--
(Structurally, one might view this state as being glassy.
%--
But, in contrast with a glass, the crosslink-rooted origins of the loss of liquidity are evident, extrinsic rather than collectively self-generated, and essentially permanent.)\thinspace\
%--
Concomitantly, from the macroscopic viewpoint the emergent characteristic of the state that arises upon sufficient crosslinking is that it responds to a static shearing stress by developing a time-persistent shear strain (and not strain {\it rate\/}, as a fluid would). It deforms, but only so far. It has become an elastic solid.

What avenues are open to the physicist concerned with developing a theory of the solid that the vulcanization process gives rise to?
%--
In view of the wide separation between the extremely long timescale associated with the breaking of the crosslinks and the comparatively short timescale for relaxational motions of the remaining, unconstrained, microscopic degrees of freedom, vulcanized matter can be regarded as possessing states of thermodynamic equilibrium in which the relaxational motions (\ie, the annealing degrees of freedom) have equilibrated in the presence of the (essentially permanent) random constraints presented by the crosslinks (\ie, the quenched variables, which persist and do not equilibrate -- see Ref.~\cite{ref:knotted} for matters of topology).
%--
Thus, from microscopic, macroscopic and conceptual perspectives, the solid phase exhibited by sufficiently crosslinked matter warrants the name {\it equilibrium amorphous solid\/}.
%--
From the theorist's standpoint, the wide separation between relaxational and crosslink-breaking timescales is precisely what is required to sanction addressing crosslinked matter \via\ the techniques of equilibrium statistical mechanics (applied to the unconstrained degrees of freedom) in the presence of quenched randomness (\ie, the information describing the random constraints on motions imposed by the crosslinks).

Having established that there is equilibrium physics to be understood, choices about the level of description remain.
%--
One may envision a truly microscopic approach, in which chemically realistic models of the molecules and crosslinking bonds are retained.
%--
Or one may envision a less specific, semi-microscopic approach, in which the molecular configurations and crosslinks are idealized as randomly fluctuating, self-avoiding quasi-Brownian curves in space, subject to the constraints that the crosslinking sites on the curves be collocated in space.
%--
This was the approach taken by Deam and Edwards some fifty years ago, in their bold, definitive formulation of  statistical mechanics in the presence of random crosslinks~\cite{D+Ephtl-1976}.

The level that we choose to focus on here for addressing soft random solids is that of statistical field theory.
%--
At this level, one is even less specific, forgoing the details of any particular model system composed of prescribed degrees of freedom (although it undoubtedly remains useful to keep the Deam-Edwards picture of the microscopics in mind and echoes of it will remain throughout this review).
%--
Instead, one focuses on identifying an appropriate continuum field degree of freedom (sometimes referred to as a collective coordinate) and gleaning from the microscopic picture the physical interpretation of the field and its mathematical transformation properties under the system's symmetries.
%--
One also focuses on identifying the minimal form of the statistical weight to ascribe to field configurations, for use in functionally averaging over the field.
%--
Statistical moments of the field then furnish coarse-grained characterizations of the equilibrium states of the system: the average of the field determines the structure of the state, including its symmetry; its fluctuation correlations determine relative spatial organization and, thus, how the system responds to external agents.
%--
So far, this all follows the Landau-Wilson vision.
%--
But, how it plays out for the statistical field theory of vulcanized matter is quite surprising, as we shall see, most notably because statistical information about the spatial heterogeneity of the amorphous solid state turns out to be encoded in a subtle manner \via\ the wave-vector dependencies of the average field and its fluctuation correlations.

%------ POSSIBLY OMIT THIS SECTION AND USE IT ELSEWHERE / BEGIN -----------------
To illustrate the microscopics/fields dyad, consider the familiar system of Ising spins populating the sites of a cubic lattice in three spatial dimensions, all interacting with their nearest neighbors \via\ a uniform coupling.
%--
In statistical field theory, one exchanges this microscopic picture of spins on a lattice for continuum theory of a real scalar field, one exchanges summing over the configurations of the spins for functionally integrating over the configurations of the fields, and one analyzes field expectation values instead of spin ones.
%--
One can of course construct the field theory from the microscopic model and match correlation functions on either side (see, e.g., Ref.~\cite{BLZ-DG}).
%--
However, one can also directly argue for the field theory on the basis of symmetries, length-scales, and dimensionalities.
%--
Furthermore, the field theory has greater generality than any particular model -- many microscopic models lead to the same field theory -- and, if one's primary aim is to identify and analyze properties that are shared across classes of systems then the direct field-theory approach, fueled by renormalization-group ideas, is particularly effective.
%------ POSSIBLY OMIT THIS SECTION AND USE IT ELSEWHERE / END -----------------

In many settings it is straightforward to identify the appropriate fields, their physical meaning, and the independent variables on which they depend.
%--
Similarly, determining the form of any constraints on the fields typically presents few subtleties.
%--
For vulcanized matter and the equilibrium amorphous solid state it exhibits, however, we shall see that the choice of field -- what values it takes, what is depends on, what constraints it obeys and, especially, its information-encoding ability -- has important nuances to it.
%--
In fact, the field embodies the statistical reflection of the intrinsically random character of vulcanized matter and the amorphous solid state.
%--
In hindsight at least, this seems natural and perhaps even obligatory.

%--
\mysecskip
\noindent{\bf What field describes amorphous solidification?\/}:
% \label{sec:whatfields}
%--
We now review the basis for a theory capable of delivering the universal properties of vulcanized matter in the liquid state (exhibited when the crosslink density is smaller than a certain threshold value), in the amorphous solid state (when the crosslink density exceeds that threshold), and in the critical state that separates the two.
%--
The essential parameters are:
    (i)~the dimension $D$ of the space inhabited by the system;
    (ii)~a lengthscale $\xi_{0}$ characterizing the size of the objects being
    crosslinked to one another, which is the analog of the lattice spacing in a spin system
    (for polymer systems we may take this to be the radius of
    gyration of the molecule in the uncrosslinked liquid state); and
    (iii)~the density of crosslinks $\crlide$ (or some proxy for it, such as $\ctp\sim{\crlide}-{\crlide_{\rm cr}}\,$, that vanishes at the transition)
    which plays a role analogous to that played by the spin-spin interaction strength
    divided by the temperature in a temperature-driven magnetic phase transition.

Now for the central matter of choosing the field, noting that its expectation value should serve as a detector and diagnostic of the amorphous solid state, with more detailed diagnostics being decodable from the the field's fluctuation correlations.
%--
With this in mind, let us ask: what is there to detect?

To answer this question, we revert to the semi-microscopic viewpoint.
%--
The liquid state arises when the system consists of a heterogeneous assembly of finite-sized components, each built from one or more molecules crosslinked together.
%--
The state is more or less viscous -- a matter of dynamics -- but every component exercises the option of wandering throughout the container, given sufficient time.
%--
No component is localized, in the sense of undergoing position-fluctuations that are only finite in range.
%--
More than that, the state has full translational symmetry.
%--
As a result of this symmetry, if we consider a system of $N$ degrees of freedom in a thermodynamically large, $D$-dimensional, cubic box of volume $V$, on which periodic boundary conditions are imposed, and if we denote equilibrium (\ie, thermal) expectation values \via\ $\langle\cdots\rangle$, then in the liquid state we have:
\begin{equation}
\label{eq:liquid}
\langle\delta\left({\bf r}-{\bf R}_{j}\right)\rangle={V}^{-1}
\,\,\,{\rm or,\,\, equivalently,}\,\,\,
\langle\exp i{\bf k}\cdot{\bf R}_{j}\rangle=\delta_{{\bf k},{\bf 0}}
\quad
({\rm for}\,\, j=1,\ldots,N).
\end{equation}
%--
Here,
${\bf r}$ is an arbitrary $D$-dimensional position-vector in the cubic box,
${\bf k}$ is an arbitrary $D$-dimensional wave-vector having components quantized by the boundary conditions to integer multiples of $2\pi/V^{1/D}$, and
$\left\{{\bf R}_{j}\right\}_{j=1}^{N}$ are the thermally fluctuating, semi-microscopic, spatial $D$-vector locations of the underlying molecular degrees of freedom.
%--
For polymers, the $\left\{{\bf R}_{j}\right\}_{j=1}^{N}$ would be the positions of each of the independent segments on each polymer chain.
%--
The expectation values in Eqs.~(\ref{eq:liquid}) are consistent with the requirements of translational invariance, under which
$\left\{{\bf R}_{j}\right\}_{j=1}^{N}\to
 \left\{{\bf R}_{j}+{\bf a}\right\}_{j=1}^{N}$
 (for any ${\bf a}$), so that, \eg,
$\langle\exp i{\bf k}\cdot{\bf R}_{j}\rangle\to
 \langle\exp i{\bf k}\cdot{\bf R}_{j}\rangle
 \exp i{\bf k}\cdot{\bf a}\,$.
%--

As for the solid state, it arises when at least some fraction of the degrees of freedom no longer wander uniformly throughout the container.
%--
We characterize the degrees of freedom in this fraction as being {\it localized\/}; when $j$ refers to a localized degree of freedom we have, instead of Eqs.~(\ref{eq:liquid}):
\begin{equation}
\label{eq:solid}
\langle\delta\left({\bf r}-{\bf R}_{j}\right)\rangle
\approx
\left(2\pi\xi_{j}^{2}\right)^{-\frac{D}{2}}
% \exp\left(-\vert{{\bf r}-{\boldsymbol\mu}_{j}}\vert^{2}/2\xi_{j}^{2}\right)
{\rm e}^{-\vert{{\bf r}-{\boldsymbol\mu}_{j}}\vert^{2}/2\,\xi_{j}^{2}}
\ne{V}^{-1}
% \quad{\rm or}\quad
  \,\,\,{\rm or}\,\,\,
\langle\exp i{\bf k}\cdot{\bf R}_{j}\rangle\approx
% \exp\left(-\xi_{j}^{2}\vert{\bf k}\vert^{2}/2\right)
{\rm e}^{i{\bf k}\cdot{\boldsymbol\mu}_{j}-\xi_{j}^{2}\,\vert{\bf k}\vert^{2}/2}
\ne\delta_{{\bf k},{\bf 0}}\,.
\end{equation}
Here, ${\boldsymbol\mu}_{j}$ is the mean position of degree-of-freedom $j$, and $\xi_{j}$ (which we refer to as its localization length) is finite and measures the RMS thermal displacement from ${\boldsymbol\mu}_{j}$.
%--
We call the expectation values~\eqref{eq:solid} the {\it density fluctuation\/} of degree-of-freedom $j$.
%--
Of course, the spatial profiles need not be Gaussian, nor need the localization be isotropic, so at this stage the profiles should be regarded as illustrative rather than necessary. (Later, we shall see that these forms are in fact adequate, at least near the solidification transition.)\thinspace\
%--
Regardless of the degree of crosslinking, the laws that govern the motion of the system (including the spatial boundary conditions, which are periodic) are invariant under translations.
%--
No longer invariant under translations, the expectation values in Eqs.~(\ref{eq:solid}) signal the spontaneous breakdown of translational symmetry.
However, as we discuss in detail below, in contrast with the case of crystallization,
$\{{\boldsymbol\mu}_{j},\xi_{j}\}_{j=1}^{N}$ are {\it random\/} consequences of the crosslinking.
%--
In particular, in view of the amorphous spatial arrangement of the mean positions of the localized degrees of freedom $\{{\boldsymbol\mu}_{j}\}_{j=1}^{N}$, whatever the wave-vector ${\bf k}$ (except ${\bf 0}$),
% the {\it particle-wide average\/}
  the {\it degree-of-freedom-wide average\/}
of the density fluctuations vanishes, \ie,
\begin{equation}
\label{eq:looksliquid}
\frac{1}{N}\sum_{j=1}^{N}
\big\langle{\rm e}^{i{\bf k}\cdot{\bf R}_{j}}\big\rangle=
\frac{1}{N}\sum_{j=1}^{N}
% {\rm e}^{i{\bf k}\cdot{\boldsymbol\mu}_{j}-\xi_{j}^{2}\vert{\bf k}\vert^{2}/2}=0
\exp\big({i{\bf k}\cdot{\boldsymbol\mu}_{j}-\tfrac{1}{2}\xi_{j}^{2}\vert{\bf k}\vert^{2}}\big)=0
\quad
({\rm for}\,\,{\bf k}\ne{\bf 0})\,.
\end{equation}
%--
Evidently, taken as a whole this quantity does not discriminate between the liquid and amorphous solid phases, even though the individual terms in the summations do, so it does not serve as an order parameter for the amorphous solid state.
%--
Said another way, the amorphous solid state is translationally invariant {\it macroscopically\/} but not {\it microscopically\/}.
%--
The amorphousness of the $\{{\boldsymbol\mu}_{j}\}_{j=1}^{N}$ obscures the symmetry breakdown, so detecting it requires a more refined approach.
%--
But despite the absence of any structure-generating, crystalline motif that would fix the mean positions of far-away molecules, the amorphous solid does possess a kind of ordering that makes sufficiently vulcanized matter qualitatively distinct from its parent liquid.
%--
All pairs of localized particles permanently retain \lq\lq knowledge\rq\rq\ of their separations blurred by thermal fluctuations, so tomorrow's instantaneous molecular organization, although partially random, is at least correlated with today's.
%--
Because of this, a remedy exists for detecting when the system has undergone a transition to the solid phase, even though the structure (\ie, the mean positions of the molecules) has no long-ranged crystallinity~\cite{ref:ReLaVe}.

What is the remedy? The obvious idea, which is on the right track but not fully sufficient, is to apply the strategy introduced Edwards-Anderson in the setting of spin glasses~\cite{E+Ajopl-1975}, and consider, instead of Eq.~(\ref{eq:looksliquid}), the second moment of the density fluctuations:
\begin{equation}
\label{eq:twoEA}
\frac{1}{N}\sum_{j=1}^{N}
\big\langle{\rm e}^{i{\bf k}^{1}\cdot{\bf R}_{j}}\big\rangle
\big\langle{\rm e}^{i{\bf k}^{2}\cdot{\bf R}_{j}}\big\rangle=
\frac{1}{N}\sum_{j=1}^{N}
{\rm e}^{i({\bf k}^{1}+{\bf k}^{2})\cdot{\boldsymbol\mu}_{j}
-(\vert{\bf k}^{1}\vert^{2}
+\vert{\bf k}^{2}\vert^{2})\,\xi_{j}^{2}/2}
\quad
({\rm for}\,\,{\bf k}^{1},{\bf k}^{2}\ne{\bf 0})\,.
\end{equation}
%--
This meets the objective of finding a quantity that discriminates between the liquid, amorphous solid (and, if necessary, crystalline phases), as it vanishes for the liquid, but for the amorphous solid it is nonzero as long as one selects
% ${\bf k}^{1}+{\bf k}^{2}={\bf 0}$,
$ ({\bf k}^{1},{\bf k}^{2})=({\bf k},-{\bf k})$,
so that
\begin{equation}
\label{eq:twoEAatZero}
\frac{1}{N}\sum_{j=1}^{N}
\big\langle{\rm e}^{ i{\bf k}\cdot{\bf R}_{j}}\big\rangle
\big\langle{\rm e}^{-i{\bf k}\cdot{\bf R}_{j}}\big\rangle=
\frac{1}{N}\sum_{j=1}^{N}
{\rm e}^{-\vert{\bf k}\vert^{2}\,\xi_{j}^{2}}=
(1-Q)\,\delta_{{\bf k},{\bf 0}}+
Q\int{d\xi}\,{\cal P}(\xi)\,{\rm e}^{-\vert{\bf k}\vert^{2}\,\xi^{2}}.
\end{equation}
Here, we have introduced $Q$ and $(1-Q)$, which are, respectively, the fraction of localized and unlocalized degrees of freedom as well as the normalized distribution
${\cal P}(\xi)\propto\sum_{j, {\rm loc.}}\delta(\xi-\xi_{j})$
of localization lengths $\xi$.
%--
Observe the singular limit: at ${\bf k}={\bf 0}$, Eq.~\eqref{eq:twoEAatZero} is trivial, but the limit ${\bf k}\to{\bf 0}$ (by which we mean $\vert{\bf k}\vert^{-1}$ much larger than the largest localization length) identifies the fraction of localized degrees of freedom.
%--
Observe, too, the emergence of a {\it statistical characterization\/} of the amorphous solid state, furnished here by the distribution ${\cal P}$. We see that the quantity introduced to discriminate between the liquid and amorphous solid state does more than that.
%--
Its wave-vector dependence encodes the heterogeneity in the localization of the system's localized degrees of freedom \via\ its Laplace-type transform from $\xi$ to $\vert{\bf k}\vert^{2}$.
%--
This is our first example of the subtle encoding of heterogeneity information touched upon, above.

But why is the second moment of Eq.~(\ref{eq:twoEA}) not fully sufficient?
%--
The reason is that the density fluctuation factors
$\langle\exp i{\bf k}\cdot{\bf R}_{j}\rangle$
are not necessarily small.
%--
Their wave-vectors are tunable continuously (except for finite-volume quantization) to values much smaller than $1/\xi_{j}\,$, in which case the magnitudes of the factors tend to unity.
%--
Thus, all moments of them matter in equal measure, and we must therefore contend with arbitrary moments,
\begin{equation}
\label{eq:AEA}
\frac{1}{N}\sum_{j=1}^{N}
\prod_{a=1}^{A}
% \langle{\rm e}^{i{\bf k}^{a}\cdot{\bf R}_{j}}\rangle=
  \big\langle\exp{i{\bf k}^{a}\cdot{\bf R}_{j}}\big\rangle=
\frac{1}{N}\sum_{j=1}^{N}
\exp
\Big(
{i{\boldsymbol\mu}_{j}\cdot\sum_{a=1}^{A}{\bf k}^{a}
-\tfrac{1}{2}\xi_{j}^{2}\sum_{a=1}^{A}\vert{\bf k}^{a}\vert^{2}}
\Big)
% {\rm e}^{i{\boldsymbol\mu}_{j}\cdot\sum_{a=1}^{A}{\bf k}^{a}
% -\frac{1}{2}\xi_{j}^{2}\sum_{a=1}^{A}\vert{\bf k}^{a}\vert^{2}},
\end{equation}
with at least two of the $A$ ($\ge 2$) wave-vectors nonzero.
%--
Smallness of this quantity, if it occurs, happens not because the moments are small but because the fraction of nonzero contributions to the summation over degrees of freedom is small, \ie, the localized fraction $Q$ is small.

We are closing in on the field out of which to build a statistical field theory.
%--
To complete the task, we appeal to the semi-microscopic formulation of Deam and Edwards~\cite{D+Ephtl-1976}, which brings up two key factors.
%--
First, two ensure that the annealed degrees of freedom are treated as equilibrating in the presence of a fixed set of quenched constraints, which are then appropriately averaged over, Deam and Edwards invoke the replica technique.
%--
This suggests that if the field is to represent Eq.~(\ref{eq:AEA}) -- or, better still, its average over realizations of the crosslinking -- then the field should depend on $n$ wave-vectors and the limit $n\to 0$ should be taken.
%--
Second, if realizations of the quenched randomness are to be averaged over, then a physically motivated measure for their probabilities should be used.
%--
Deam and Edwards proposed a masterful resolution of this issue, arguing that the likelihood of finding some particular set of crosslinks is proportional to the likelihood of finding the associated set of degrees of freedom to be in contact with one another {\it in the uncrosslinked, parent liquid\/}.
%--
A little reflection reveals that this likelihood is proportional to the partition function of the system when it is subject to precisely those crosslink constraints, so the sought measure may be unknown -- but it is a {\it known\/} unknown.

The upshot, then, is that the appropriate field $\Omega$ depends on $1+n$ replicas of ${\bf k}$ (or its Fourier conjugate, ${\bf r}$), so it reads
$\Omega({\bf k}^{0},{\bf k}^{1},\ldots,{\bf k}^{n})$ or
$\Omega({\bf r}^{0},{\bf r}^{1},\ldots,{\bf r}^{n})$.
%--
(Within the Deam-Edwards framework, it serves as the Hubbard-Stratonovich field or collective coordinate for decoupling the interaction term that originates in the crosslinking constraints; see Refs.~\cite{B+E-1980,GGprl-1987,CGZepl-1994,GCZaip-1996})\thinspace\
%--
Its field-theoretic average $\langle\Omega\rangle$, computed using a weight that we shall address shortly, is {\it the order parameter for amorphous solidification\/}.
%--
The physical meaning is the extention to the replica theory of Eq.~(\ref{eq:AEA}), and is given by
\begin{equation}
\label{eq:means}
\big\langle\Omega({\bf k}^{0},{\bf k}^{1},\ldots,{\bf k}^{n})\big\rangle
\big\vert_{{\bf k}^{0}={\bf 0}}=
\Big[\,
\frac{1}{N}\sum_{j=1}^{N}
\prod_{\alpha=1}^{n}
\big\langle{\rm e}^{i{\bf k}^{\alpha}\cdot{\bf R}_{j}}\big\rangle
\Big],
% ({\rm for}\,\,{\bf k}^{1},{\bf k}^{2}\ne{\bf 0})\,,
\end{equation}
%--
in which the brackets $[\cdots]$ indicate averaging over the realizations of the crosslinking.
%--
Whilst, as we shall see, all $1+n$ replicas feature in an equivalent way in the field theory's weight functional, the zeroth replica (which arrived to generate the measure for the quenched randomness) has a distinct meaning, as it probes the auxiliary ensemble that generates the quenched randomness measure. Thus, the moments of the physical density fluctuations are associated with replicas 1 to $n$. That is why we have set the wave-vector ${\bf k}^{0}$ to zero in exhibiting the physical meaning of the order parameter in Eq.~\eqref{eq:means}.

To precisely settle the issue of the field degrees of freedom needed to address the critical phenomenon of amorphous solidification, the last ingredient is a set of inconsequential-looking (but in fact essential) constraints:
\begin{subequations}
\begin{equation}
\label{eq:constraint}
\int_{V}\prod_{\alpha=0}^{n}d^{D}r^{\alpha}\,
\Omega({\bf r}^{0},\ldots,{\bf r}^{n})\!=\!1,
% \,\,{\rm and}\,\,
\,
\int_{V}\prod_{\alpha=0}^{n}d^{D}r^{\alpha}\,
\Omega({\bf r}^{0},\ldots,{\bf r}^{n})
{\rm e}^{i{\bf q}\cdot{\bf r}^{\beta}}
\!=\!0
\,\,
({\bf q}\ne{\bf 0};\beta\!=\!0,\ldots,n),
\end{equation}
which have an even simpler form in wave-vector space, \viz,
\begin{equation}
\label{eq:constraint}
\Omega({\bf k}^{0},\ldots,{\bf k}^{n})
\big\vert_{\{{\bf k}^{{\alpha}}={\bf 0}\}_{{\alpha}=0}^{n}}=1,
% \,\,{\rm and}\,\,
\,\,
\Omega({\bf k}^{0},\ldots,{\bf k}^{n})
\big\vert_{\{{\bf k}^{\alpha}={\bf q}\,
\delta^{\alpha,\beta}\}_{{\alpha}=0}^{n}}=0
\,\,\,
({\bf q}\ne{\bf 0};\beta=0,\ldots,n).
\end{equation}%
\label{eq:together}%
\end{subequations}%
%--
The first of these two constraints simply says that the number of particles remains at $N$ and does not fluctuate.
%--
The latter -- a family labelled by $\beta$ and a nonzero wave-vector ${\bf q}$ -- says that macroscopic density waves (such as those that would acquire a nonzero expectation value if a periodic crystal were to form) remain at zero amplitude and do not fluctuate.
%--
It incorporates the physical features that:
(i)~inter-particle repulsions stabilize the spatially uniform liquid; and
(ii)~the associated degrees of freedom remain stabilized by the inter-particle repulsions, even when other degrees of freedom, including those associated with amorphous solid ordering, become destabilized as a result of crosslinking, and the (unstable) liquid equilibrium state is exchanged for the (stable) equilibrium amorphous solid state.
%--
(Alternatively, one could opt to relax these constraints, and instead heavily penalize the associated excitations.)\thinspace\

It is useful to introduce some nomenclature to identify the constrained degrees of freedom.
%--
We refer to them as spanning the {\it lower replica sector\/} of the field $\Omega$ (or LRS).
We refer to the complementary part of the field as the {\it higher replica sector\/} of $\Omega$ (or HRS).
%--
Fields for which two or more replicated wave-vectors are nonzero (\ie, the HRS) are left unconstrained, and are the essential field degrees of freedom for describing amorphous solidification.
%--
It is tempting to assume that the LRS degrees of freedom, residing as they do at the origin or on the axes of replicated wave-vector space, would constitute an infinitesimal fraction of the field degrees of freedom in the macroscopic-system limit, so that their removal would be of little consequence.
%--
But for the replica limit, this would be true; but their removal is in fact a {\it sine qua non\/} for obtaining physical results.

%--
\mysecskip
\noindent{\bf Which statistical field theory describes amorphous solidification?\/}:
% \label{sec:whatfieldtheory}
%--
Having identified the field $\Omega$ as being suitable for capturing the physical properties of the equilibrium amorphous solid state and the crosslink-triggered transition to it, we now turn to the question of developing a minimal model that governs the equilibrium value and fluctuations of $\Omega$; see Refs.~\cite{PCGZprb-1998,HECphd-1998,ref:Peng-2000}.
%--
Since our focus is on the solidification transition regime, we seek a Landau-Wilson Hamiltonian functional $\efe[\Omega]$ that provides a suitable weight, $\exp(-\efe)$, for field configurations.
%--
On general grounds, the terms that we expect to need involve limited numbers of powers of $\Omega$ and its gradients.
%--
The terms must also be invariant under {\it independent\/} translations (and rotations) of the replicas, \ie, $\efe[\Omega^{\prime}]=\efe[\Omega]$ for
$\Omega^{\prime}({\bf k}^{0},\ldots,{\bf k}^{n})=
 \exp(i\sum\nolimits_{\alpha=0}^{n}{\bf k}^{\alpha}\cdot{\bf a}^{\alpha})\,
 \Omega({\bf k}^{0},\ldots,{\bf k}^{n})$,
where the vectors $\{{\bf a}^{\alpha}\}_{\alpha=0}^{n}$
are the $1+n$ independent translations of the replicas.
%--
(Rotations have corresponding transformation laws, but it is not necessary to focus on them.)\thinspace\
%--
At the transition, this symmetry of independent translations of the replicas spontaneously breaks, leaving as the residual symmetry the {\it common\/} translations of the replicas, \ie, $({\bf a}^{0},\ldots,{\bf a}^{n})=({\bf a},\ldots,{\bf a})$. This mechanism marks the emergence of a state in which some fraction of the microscopic degrees of freedom have become localized, but randomly, in the sense that their mean positions possess no macroscopic periodicity.
%--
Thus, we arrive at the Landau-Wilson form (see Ref.~\cite{PCGZprb-1998}):
%--
% % \begin{equation}
% \efe=\pno\sum_{\khat\in\hrs}\Big(
% -\frac{a}{2}\ctp+  \frac{\barexi^{2}}{2}
% \khat^{2}\Big)\vert\opf(\khat)\vert^{2}-\pno\cocon
% \!\!\!\sum_{\khat_{1},\khat_{2},\khat_{3}\in\hrs}\!\!\!
% \delta_{\zhat,\khat_{1}+\khat_{2}+\khat_{3}}\,
% \opf(\khat_{1})\,\opf(\khat_{2})\,\opf(\khat_{3}),
% \label{eq:Landau-Wilson}
% \end{equation}
%--
\begin{equation}
\pno^{-1}\efe=\sum_{\khat\in\hrs}
\big(-\ctp+ \tfrac{1}{2}\khat^{2}\big)
\vert\opf(\khat)\vert^{2}-
\!\!\!\sum_{\khat_{1},\khat_{2},\khat_{3}\in\hrs}\!\!\!
\delta_{\zhat,\khat_{1}+\khat_{2}+\khat_{3}}\,
\opf(\khat_{1})\,\opf(\khat_{2})\,\opf(\khat_{3}),
\label{eq:Landau-Wilson}
\end{equation}%
%--
% [rescalings of the effective Hamiltonian,
% the control parameter,
% and the unit of length;
% Cf Castillo thesis, p. 88]
%--
in which we have introduced the convenient notation
$\khat\equiv({\bf k}^{0},\ldots,{\bf k}^{n})$
for replicated wave-vectors and
$\khat^{2}\equiv\sum\nolimits_{\alpha=0}^{n}\vert{\bf k}^{\alpha}\vert^{2}$.
%--
In this Hamiltonian, $\ctp$ is the dimensionless control parameter for the transition and is determined, \eg, by the density of crosslinks.
%--
If derived from the semi-microscopic theory, it would come with an additional factor $a/2$ that characterizes the critical value of the likelihood of crosslinking formation.
%--
Similarly, the gradient-term factor $\khat^{2}$ would come with an additional factor $\barexi^{2}$, which is a (squared) length characterizing the bare size of the units being connected and which serves as an effective (squared) lattice spacing.
The cubic term would come with a factor $\cocon$, which measures the strength of the interactions between field fluctuations.
%--
Convenient rescalings allow us to set $a/2$, $\barexi$ and $g$ to unity.
%--
In addition, we measure energies in units of the thermal energy scale $k_{\rm B}T$, where $T$ is the temperature.
%--

Some observations on the form of $\efe$:
(i)~Only the HRS degrees of freedom feature in it.
%--
They constitute the set of critical degrees of freedom for the amorphous solidification transition.
%--
One must be careful to exclude the LRS degrees of freedom when performing summations over replicated wave-vectors.
%--
(ii)~The theory is cubic. At first sight, this suggests that the transition is discontinuous.
%--
However, the signs of the terms in Eq.~\eqref{eq:Landau-Wilson} are such that the discontinuous transition would be to a state marked by a negative fraction of localized degrees of freedom, which is therefore unphysical.
%--
It is thus appropriate to limit our attention to the physical branch, and for that branch the transition is continuous.

%--
% \subsection{Equilibrium state\label{sec:equilib}}
%--
\mysecskip
\noindent{\bf What is the structure of the equilibrium state?\/}:
%--
We now address $\langle\opf\rangle$ (\ie, the equilibrium value of the amorphous solid order parameter) within the mean-field (or classical) level of approximation (\ie, neglecting field fluctuations).
%--
In this case, $\langle\opf\rangle$ is given by the value of $\opf$ (which we call $\opfe$) that makes $\efe$ stationary (see Refs.~\cite{CGZepl-1994,GCZaip-1996,HECphd-1998,PCGZprb-1998}):
\begin{equation}
0={\delta\efe\over{\delta\opf(-\khat)}}\Bigg\vert_{\opfe}=
2\big(
-\ctp+\tfrac{1}{2}\khat^{2}
\big)\,\opfe(\khat)-
3\!\!\!\sum_{\khat_{1},\khat_{2}\in\hrs}\!\!\!
\delta_{\khat,\khat_{1}+\khat_{2}}
\,\opfe(\khat_{1})\,\opfe(\khat_{2}).
\label{eq:statcon}
\end{equation}%
For $\ctp\le 0$, the liquid state, $\opfe=0$, is the stable minimizer of $\efe$.
%--
For $\ctp>0$, the liquid state loses its stability and is replaced by a stable minimizer that has precisely the amorphous solid state form anticipated in Eq.~\eqref{eq:AEA}:
\begin{equation}
\opf(\khat)=
\locfrac\,\delta_{\zvec,\tilde{\kvec}}
% \int\! dt\,\dist(t)\,{\rm e}^{-\khat^{2}/2t},
  \int\! d\xi\,\dist(\xi)\,{\rm e}^{-\khat^{2}\xi^{2}/2},
\label{eq:opform}
\end{equation}%
% where $t$, an inverse square localization length,
% is integrated over all positive, finite values
where $\xi$, a localization length,
is integrated over all positive, finite values
(so as to feature only localized particles),
$\tilde{\kvec}\equiv\sum_{\alpha=0}^{n}\kvec^{\alpha}$ and, again,
$\khat^{2}\equiv\sum_{\alpha=0}^{n}\kvec^{\alpha}\cdot\kvec^{\alpha}$.
%--
The diagnostic $\locfrac$ has the meaning of the fraction of localized particles.
%--
$\dist$ is the normalized distribution of localization lengths.
%--
It is a characterization of the heterogeneity of the state that, as we now see, is encoded in the equilibrium value of the order parameter.
%--
For a detailed analysis of the linear stability of the amorphous solid state, see Ref.~\cite{HECphd-1998,CGZprb-1999}.

Computable (\ie, replica-free) equations for $\locfrac$ and $\dist$ follow from inserting the form~\eqref{eq:opform} into the stationarity condition~\eqref{eq:statcon}.
%--
In this way, one arrives at the mean-field self-consistency conditions:
\begin{subequations}
\begin{eqnarray}
\label{eq:perc}
0&=&
-2\,\ctp\,\locfrac+3\,\locfrac^{2},\\[4pt]
0&=&
{\scat^{2}\over{2}}{d\scadist\over{d\scat}}
-(1-\scat)\,\scadist(\scat)+(\scadist\circ\scadist)(\scat),
\label{eq:sceforscadist}%
\end{eqnarray}%
\label{eq:class-state}%
\end{subequations}%
in which $t$ and $\dist$ have been exchanged for their scaled forms ($\scat$ and $\scadist$), defined in terms of the control parameter $\ctp$ \via\
\begin{subequations}
\begin{eqnarray}
% t&=&(\ctp/2)\,\scat,\\
  \xi^{-2}&=&(\ctp/2)\,\scat,\\
% \dist(t)
  ({2/{\xi^{3}}})\,\dist(\xi)
&=&(\ctp/2)^{-1}\,\scadist(\scat),
\end{eqnarray}%
\end{subequations}%
and where $\scadist\circ\scadist$ indicates the Laplace convolution of $\scadist$ with itself.
%--
Focusing on Eq.~\eqref{eq:perc}, we see that the liquid state is associated with the solution $\locfrac=0$ (valid for $\ctp\le 0)$, whilst the amorphous solid state is associated with the solution $\locfrac=2\,\ctp/3$ (which holds for $\ctp>0$), indicating the emergence of a localized fraction of particles.
%--
As for $\dist$, its asymptotics have been obtained for small and large $\xi$, and its specific form has been determined numerically for all $\xi$; see Refs.~\cite{CGZepl-1994,GCZaip-1996}.
%--
Qualitatively, $\dist$ peaks at $\xi$ of order $\ctp^{-1/2}$ and tends rapidly to zero for larger and smaller values of $\xi$.
%--
Quantitative support for these predictions, including data collapse on to the universal scaling form for the distribution of localization lengths, was found in numerical simulations by Barsky and Plischke~\cite{refs:SFU-group} and in experiments on colloidal and protein gels by Dinsmore and Guertin (unpublished).

Let us briefly reflect on the status of the results that this mean-field theory has delivered.
%--
(i)~The fraction of localized particles vanishes as the transition is approached as $Q\sim\ctp^{\beta}$ with $\beta=1$.
%--
Appropriately, this exponent is consistent with the mean-field theory of percolation and also the corresponding random graph theory results of Erd{\"o}s and R{\'e}nyi (see Ref.~\cite{ref:Erdos}).
%--
(ii)~The typical localization length diverges as the transition is approach as
$\xi_{\rm typ}\sim\ctp^{-\nu}$ with $\nu=1/2$.
%--
(iii)~In addition, mean-field theory presents a single, parameter-free scaling function $\scadist$, which governs the distribution of localization lengths for all near-critical values of the crosslink density.
%--
Precisely this distribution was previously found
by Stinchcombe~\cite{ref:RBS-CBL}, in the setting of the conductivity of a disordered Bethe lattice, and
by Stephen~\cite{ref:MJS-RRN}, in the setting of a random resistor networks.

%--
\mysecskip
% \section{Beyond mean-field theory\label{sec:MFTplus}}
\noindent{\bf How can we determine critical phenomenology beyond mean-field theory?\/}:
%--
The results discussed in the previous section were obtained within mean-field theory, \ie, under the neglect of fluctuations of the field $\opf$. Thus, we should ask: under what conditions are the mean-field results strongly altered by fluctuations? The answer is determined by the appropriate Ginzburg criterion, which in this case is a condition on the crosslink density $\crlide$.
%--
Almost fifty years ago, by focusing on fluctuations in the number of gel segments within a correlation volume, De~Gennes~\cite{ref:DeGennes-1977} was able to arrive at the criterion:
\begin{equation}
\left\vert
({\crlide-{\crlide_{\rm cr}}})
/\,{\crlide_{\rm cr}}
\right\vert\alt
\left({L/{\ell}}\right)^{-(D-2)/(6-D)}
\left({\varphi/{g^{2}}}\right)^{-2/(6-D)},
% Narrower, MFT good, for long chains and high volume fraction
%--
\end{equation}
where $\crlide_{\rm cr}$ is the crosslink density at which amorphous solidification first occurs; $(L/{\ell})$ [$\,\approx{\barexi^{2}/\ell^{2}}\,$] is the number of polymer persistence lengths $\ell$ per molecule length $L$ (\ie, the number of statistically independent segments per chain); $\varphi$ is the volume fraction (\ie, a measure of how much volume is filled by polymer as opposed to solvent); $\cocon$ is the strength of the field nonlinearity; and, as we already know, $D$ is the dimension of space.
%--
The statistical field theory approach, taken here, gives the identical Ginzburg criterion; see
Ref.~\cite{ref:Peng-2000}.
%--
It says that above an upper critical dimension of 6, mean-field theory is accurate.
%--
It also says that below six dimensions fluctuations have a strong effect within a window of crosslink densities around $\crlide_{\rm cr}$ that becomes progressively wider for shorter polymer chains, smaller polymer volume fractions, and spatial dimensions increasingly far below~6.
%--
(Later, we shall see that the lower critical dimension for the transition is 2, consistent with the fact that it is a continuous symmetry that is broken at the transition -- decisively non-percolative behavior.)\thinspace\ Amorphous-solid--formers are commonly dense, three-dimensional systems of long polymers, in which case one can expect the conclusions of mean-field theory to hold well.
%--
However, this need not be the case, so the impact of fluctuations, which we now review, is by no means a mere theoretical curiosity.

To begin exploring the consequences of fluctuations, we follow Ref.~\cite{ref:Peng-2000} and promote $\efe$ of Eq.~(\ref{eq:Landau-Wilson}) from a Landau free energy to the Landau-Wilson minimal effective Hamiltonian that controls the statistical weight of configurations of the field $\opf$.
%--
Provided we attend to the constraints that define the appropriate degrees of freedom (\ie, the HRS sector of $\opf$) and the replica limit (\ie, $n\to 0$), we are well positioned to develop an analytical, renormalization-group perspective on the amorphous solidification transition that incorporates field fluctuations and, hence,
    the thermal motion of the constituents,
    the quenched random crosslinking constraints imposed on their motion, and
    particle-particle repulsions, which suppress density fluctuations and play a pivotal role in determining the appropriate degrees of freedom.
%--
Apart from the factors concerning the replica limit and the constraints on the field, the approach follows a conventional path. So, what does the renormalization-group analysis yield in the present setting?
%--
\hfil\break\noindent
(i)~It yields the two-field correlator of order-parameter fluctuations, which signals the transition to the amorphous solid state and indicates the accompanying emergence of a macroscopic connected network.
\hfil\break\noindent
(ii)~It yields the Ginzburg criterion mentioned above, and an upper critical dimension (\viz, 6), below which the Gaussian fixed point is exchanged for the Wilson-Fisher fixed point.
\hfil\break\noindent
(iii)~{\it Via\/} an expansion in powers of $\varepsilon$
(\ie, the amount by which the spatial dimension $D$ lies below 6)
it yields certain universal critical exponents characterizing the amorphous solidification transition, most notably the exponents
    $\beta$, which characterizes the vanishing of the localized fraction (\via\ $Q\sim\ctp^{\beta}$), and
    $\nu$, which characterizes the divergence of the typical localization length (\via\ $\xi_{\rm typ}\sim\ctp^{-\nu}$),
    as the transition is approached from the solid side.
%--
To first order in $\varepsilon$, the minimal model gives
$[\beta,1/\nu]=[1-(\varepsilon/7),2-(5\varepsilon/21)]$.
%--
Identical values for the exponents that characterize the physically corresponding quantities in percolation theory had been obtained previously by Harris {\it et al\/}.~\cite{ref:Harris-1975} by means of a field-theoretic analysis of the Kasteleyn-Fortuin representation of percolation (\ie, the $m\to 1$ limit of the $m$-state Potts model).

Thus, one finds that the anticipated correspondence between vulcanization and percolation, already seen at the level of mean-field theory, continues even after fluctuation effects are factored in, in the sense that physically corresponding quantities -- one from the percolation theory side and one from the amorphous solidification theory side -- are governed by identical critical exponents, at least to first order in $\varepsilon$.
%--
The relationship between the amorphous solidification and percolation universality classes was subsequently demonstrated to all orders in $\varepsilon$ in work by Janssen and Stenull~\cite{ref:Janssen-2001} (based on an analysis of the Gell-Mann--Low renormalization-group equation) and in Ref.~\cite{ref:Peng-2001} (based on an all-orders identification with the Houghton-Reeve-Wallace field-theory approach to percolation~\cite{ref:HRW-1978}).
%--
This suggests that architecture (\ie, crosslinking) is more important than thermal fluctuations for matters of connectivity, such as system-spanning network formation.

However, the physics of percolation and of amorphous solidification are by no means identical more broadly.
%--
Percolation involves one statistical ensemble; amorphous solidification involves two: one for the annealing variables and one for the quenched variables.
%--
Percolation is a statistical question; amorphous solidification is a thermodynamic one.
%--
And, most significantly, amorphous solidification has at its heart the breakdown of translational symmetry and the emergence of elasticity and heterogeneity: it certainly features percolation, but also localization and rigidity, with all three concepts manifesting themselves through independent (although, of course, inter-related) physical phenomenology.

%--
% \section{Goldstone\label{sec:Goldstone}}
\mysecskip
\noindent{\bf What is the structure of the Goldstone excitations?\/}:
%--
We have mentioned several time alrady that the transition to the amorphous solid state is marked by the spontaneous breakdown of continuous symmetries. This is to be expected, but the pattern within the replica framework is unusual: in contrast to the liquid-state, invariance of the state under {\it relative\/} translations of the replicas is absent, but invariance under {\it common\/} translations remains.
%--
This symmetry-breaking pattern differentiates amorphous solids from crystalline ones. In amorphous solids, the random localization of the particles leaves the medium macroscopically homogeneous, which is reflected in the residual invariance under common translations.
%--
Because it is {\it continuous\/} symmetries that are broken, the Goldstone circle of ideas indicates quite generally that some kind of continuum elasticity theory should emerge to provide a characterization of the low-energy excitations and response of the emergent solid state.
%--
Indeed it does, despite the unusual symmetry-breaking pattern, as we now explain.

Recall that the amorphous solid state is detected and diagnosed \via\ the equilibrium value of the order parameter $\opf$ taking the form given in Eq.~(\ref{eq:opform}).
%--
Unpacking the notation $\tilde{\kvec}$ and $\khat^{2}$ in terms of the replicated wave-vectors, as explained shortly after Eq.~(\ref{eq:opform}), one sees that the state [\ie, the form~(\ref{eq:opform})] is indeed invariant under {\it common\/} translations of the replicas, but not if the replicas are translated {\it relative\/} to one another.
%--
(Because their variance does not vanish, we term such transformations {\it dispersive\/} translations.)\thinspace\
%--
Unlike common translations, dispersive translations change the value of the order parameter $\opf$ (although not, of course, its energy $\efe[\opf]$), moving $\opf$ from one equilibrium state of the replica theory to another.
They are the exact Goldstone modes resulting from the broken symmetry.

In view of the structure of the exact Goldstone modes, we see that the low-energy excitations of the equilibrium state can be constructed \via\ the following \lq\lq phase\rq\rq\ deformation of the Kronecker-$\delta$ factor of the order parameter in Eq.~(\ref{eq:opform}), \viz,
\begin{equation}
\delta_{\zvec,\tilde{\kvec}}=
\int{d^{D}z\over{V}}\,
% {\rm e}^{i\sum_{\alpha=0}^{n}\kvec^{\alpha}\cdot\zedvec}
\exp{i\sum_{\alpha=0}^{n}\kvec^{\alpha}\cdot\zedvec}
\longrightarrow
\,\int{d^{D}z\over{V}}\,
\exp{i\sum_{\alpha=0}^{n}\kvec^{\alpha}
\cdot\zedvec}\,
\exp{i\sum_{\alpha=1}^{n}\kvec^{\alpha}
\cdot\udisvec^{\alpha}(\zedvec)},
\label{eq:howtodeform}
\end{equation}
parametrized \via\ the $n$ (not $n+1$) position-dependent, $D$-vector-valued displacement fields $\{\udisvec^{\alpha}(\zedvec)\}_{\alpha=1}^{n}\,$.
%--
It is understood that the Fourier content of these displacement fields is supported at lengthscales at least somewhat longer than the typical localization length.
%--
This ensures that the excitations are indeed low-energy ones and do not mix with deformations of the \lq\lq amplitude\rq\rq\ factor of the order parameter
[\viz, $\locfrac\int d\xi\,\dist(\xi)\,{\rm e}^{-\khat^{2}\,\xi^{2}/2}\,$],
which necessarily incorporate excitations from gapped branches.
(Thus, we may take the \lq\lq amplitude\rq\rq\ factor to be undeformed.)\thinspace\
%--
Now, as we have discussed above, strong inter-particle repulsion effectively imposes the second of the two constraints specified in Eq.~\eqref{eq:together}.
%--
Ensuring that such penalties are not incurred is achieved be requiring that the displacement fields obey the incompressibility condition:
%--
${\rm det}
\big(
\delta_{dd^{\prime}}+
\partial u_{d}^{\alpha}(\zedvec)/
\partial z_{d^{\prime}}^{\phantom\alpha}
\big)=1$, which for small displacement gradients can be approximated by
$\partial u_{d}^{\alpha}(\zedvec)/
\partial z_{d}^{\phantom\alpha}=0$.
Said equivalently, we restrict our attention to {\it pure shear deformations\/}.
%--
These deformations are reminiscent of an elasticity theory that has been conventionally replicated to account for quenched in elastic-constant and stress-field disorder. This line of reasoning was developed in detail in Refs.~\cite{MGXZepl-2007,XMMphd-2008,MGXZpre-2009}.

The structure of the exact Goldstone modes and low-energy excitations of the equilibrium state may feel more intuitive when viewed in replicated real space.
%--
There, as the Fourier transform of Eq.~\eqref{eq:opform} readily shows, the region in
$({\bf x}^{0},{\bf x}^{1},\ldots,{\bf x}^{n})$-space
where the order parameter is large in amplitude is concentrated in a tube of width given by the typical localization length around the \lq\lq body diagonal\rlap,\rq\rq\ \ie, the \lq\lq line\rq\rq\ ${\bf x}^{\alpha}={\bf z}$ (for all $\alpha$ and any ${\bf z}$).
%--
Exact Goldstone modes amount to the $nD$ independent translations of the whole tube laterally with respect to the body diagonal.
%--
Low-energy excitations amount to lateral ripples of this tube, of wavelength rather longer than the typical localization length.

The deformation we opted for in formula~\eqref{eq:howtodeform} is not generic. Rather, it amounts to a \lq\lq gauge\rq\rq\ choice that singles out the zeroth replica.
%--
This choice is convenient for calculational purposes, but there is also a physical idea behind it.
%--
The zeroth replica is present to generate the Deam-Edwards distribution of quenched disorder, whilst the other replicas represent the physical degrees of freedom that can solidify and, hence, be subjected to deformations.

\mysecskip
\noindent{\bf What are the implications of the Goldstone excitations for elasticity at macroscopic and mesoscopic lengthscales?\/}:
%--
Now that we understand the structure of the Goldstone-deformed states, we are in a position to address a range of their physical implications.
%--
The first and most salient implication concerns the key physical characteristic of the amorphous solid state, \viz, its rigidity.
%--
To understand this, take the Goldstone-deformed state, characterized by the displacement fields $\{\udisvec^{\alpha}(\zedvec)\}_{\alpha=1}^{n}$, insert it into $\efe$ given by Eq.~\eqref{eq:Landau-Wilson}, and focus on the increase in free energy $\delta\efe$ arising from the deformation of the state, for now keeping only terms to second order in displacement fields.
%--
As shown in Ref.~\cite{ref:ScaleDep-2025} (and building on several earlier analyses of elasticity~\cite{HECphd-1998,CGpre-1998,CGpre-2000}), one obtains the free-energy cost of elastic deformations, including their dependence on the deformation scale $\qvec\,$:
%--
\begin{equation}%
%--
% \delta\efe\big[\,\opfe,\udisvecF\,\big]=
 \delta\efe=
-{\pno\over{2\vol}}
\sum_{\underset{(\alpha_{1}\,\ne\,\alpha_{2})}
        {\alpha_{1},\,\alpha_{2}\,=\,1}}^{n}
{1\over{\vol}}\sum_{\qvec}\,
\sds(\qvec)\,\vert\qvec\vert^{2}\,
\udisvecF^{\alpha_{1}}(\qvec)^{\ast}\cdot
\udisvecF^{\alpha_{2}}(\qvec),
\end{equation}%
%--
where we have exchanged the displacement fields
$\{\udisvec^{\alpha}\}_{\alpha=1}^{n}$
for their Fourier transforms
$\{\udisvecF^{\alpha}\}_{\alpha=1}^{n}$,
defined \via\ the pair
\begin{eqnarray}
\udisvecF(\qvec)=
\int{d^{D}z}\,{\rm e}^{i\qvec\cdot\zedvec}\,\udisvec(\zedvec)
\quad{\rm and\/}\quad
\udisvec(\zedvec)=
{1\over{\vol}}\sum_{\qvec}{\rm e}^{-i\qvec\cdot\zedvec}\,\udisvecF(\qvec)\,.
\end{eqnarray}%
Hence, we identify $\sds(\qvec)$ as the {\it scale-dependent elastic shear modulus\/} $\sds(\qvec)$.
%--
It is conveniently expressed as
$\sds(\qvec)\equiv\sds(\bm{0})\,\Sigma(\barexi^{2}\vert\qvec\vert^{2}/\ctp)$,
\ie, in terms of its long-distance limit,
$\sds(\zvec)\equiv k_{\rm B}T\,(4\,\ctp^{3}/27)$,
along with a dimensionless scaling function $\Sigma$.
% [** Should there be a volume, say $\barexi^{3}$, in the denominator of S(0)?]
%--
% [** Micro picture of why u is a deformation. How do we now that these really are the displacement fields.]

Thus, on the one hand we have arrived at the value of the conventional elastic shear modulus, operative for displacement fields that vary only at macroscopic lengthscales, finding, in particular, that within mean-field theory it vanishes as the third power of the excess constraint density and that it is linear in the temperature, reflecting its entropic origin.
%--
On the other hand, the scaling function $\Sigma$ determines how the shear modulus diminishes as the wavelength of the displacement is reduced, from greater than to less than the typical localization length.
%--
$\Sigma$ is completely determined by the scaled distribution $\scadist$, as shown in detail in Ref.~\cite{ref:ScaleDep-2025}, decaying smoothly from 1 to zero as its argument varies from macroscopic to microscopic scales, thus revealing how the scale-dependence of the elastic shear modulus serves as a lens on the distribution of localization lengths $\dist$.

But why focus on the scale-dependent shear modulus?
%--
There are two interconnected reasons.
%--
The first is that the transition to the amorphous solid state is a continuous one, unlike the crystalline solidification, say of mercury, which is discontinuous and thus unaccompanied by any diverging lengthscale.
%--
Thus, whilst, in principle, solid mercury has a scale-dependent elastic modulus, the scale on which it begins to fall from its macroscopic value is always on the order of the atomic spacing, even near the melting temperature.
%--
In contrast, the continuity of the amorphous solidification transition is accompanied by a scale for the typical localization length that diverges as the transition is approached: this (potentially long) length sets the scale on which the shear modulus begins to fall below its macroscopic value as elastic deformations at shorter and shorter lengthscales are considered.
%--
Why the reduction? The picture to have in mind is that particles localized on lengthscales longer than the deformation lengthscale are, in effect, liquid at this scale, and thus they barely contribute to the elasticity.
%--
The second reason for focusing on the scale-dependent shear modulus is that, again unlike crystalline solids, the amorphous solid is characterized \via\ a continuous distribution of localization lengths, centered on the typical value but spreading on either side with a width that also diverges with the typical lengthscale as the transition is approached.
%--
Thus, even if one limits attention to the range of deformation lengthscales that are longer than the typical localization length, the shear modulus already has a scale dependence: as the deformation scale is tuned down from  macroscopic, the shear modulus decreases, because a smoothly increasing fraction of the particles are localized on scales longer than the deformation scale and thus contribute only weakly to the elasticity.
%--
Hence, one sees that the scale dependence of the elastic modulus serves as a probe of the localization-length distribution.

%------ POSSIBLY OMIT THIS SECTION AND USE IT ELSEWHERE / BEGIN -----------------
Formally, one could consider deformation scales that are shorter than the typical localization length, and thus relate $\Sigma$'s further progression towards zero to the scaled distribution $\scadist$, consistent with the idea that the fraction of particles localized at such scales is very small.
%--
However, this would not be a consistent application of the Goldstone framework, because to correctly analyze such deformations would necessitate the incorporation of amplitude-type modes of order-parameter deformations.
%--
(These modes, which lie outside the scope of the Goldstone framework, are the subject of ongoing work with Boli Zhou. By obtaining them, we would be able to construct the complete Gaussian propagator, and thus learn about issues such as localization-length correlations, consequences of finite sample-size, and how elastic deformations influence the distribution of localization lengths.)\thinspace\
At an even shorter scales -- the size of the individual, pre-crosslinking polymers -- the Landau-Wilson framework ceases to hold. So, even if the amplitude sector of the theory were incorporated, one would not trust the behavior of the shear modulus at such scales or its implications for the short-length asymptotics of the distribution of localization lengths.
%------ POSSIBLY OMIT THIS SECTION AND USE IT ELSEWHERE / END -----------------

%--
% \label{sec:HOdecoding}
\mysecskip
\noindent{\bf How are mesoscale heterogeneity and its spatial correlations encoded?\/}:
%--
A recurring theme here has been that the random architectures of amorphous-solid--forming media give rise to systems whose particles exhibit (ensemble or time)-averaged thermal motions that are heterogeneous, varying randomly in character from point to point across the medium.
%--
For example, we have seen in Eq.~\eqref{eq:opform} that coded into the wave-vector dependence of the order parameter $\langle\opf\rangle$ is the simplest of the state's heterogeneity characterizations: the statistical distribution $\dist(\xi)$ governing the point-to-pont variation of particle localization lengths $\xi$.

As developed in detail in Ref.~\cite{ref:UMH}, building on earlier work (see Refs.~\cite{ref:MGZ-2004,ref:GMZ-2004}), the correlations amongst the fields of the replica field theory, such as $\langle\opf\opf\rangle$, supply increasingly refined statistical information about the heterogeneity of the thermal motions of the particles in the medium, beyond that supplied by the order parameter.
%--
The two-field correlation function, \eg, supplies a joint probability distribution $\ProbCharsLarge$, which we refer to as the {\it heterogeneity distribution\/} and which is defined as follows:
\begin{eqnarray}
\label{eq:heteroprob}
&&
% \ProbCharsLarge(\MUvecDUM_{1},\MUvecDUM_{2},\rmsFlDUM_{1},\rmsFlDUM_{2},\rmsCoDUM)
  \ProbCharsLarge(\MUvecDUM_{1}-\MUvecDUM_{2},\rmsFlDUM_{1},\rmsFlDUM_{2},\rmsCoDUM)
\\
\noalign{\smallskip}
&&\quad\equiv
\bigg[{1\over{\pno}^{2}\locfrac^{2}}
\sum_{j_{1},\,j_{2}\in\loc}
\delta(\MUvecDUM_{1}-\MUvec_{j_{1}})\,
\delta(\MUvecDUM_{2}-\MUvec_{j_{2}})\,
\delta(\rmsFlDUM_{1}-\rmsFl_{j_{1}})\,
\delta(\rmsFlDUM_{2}-\rmsFl_{j_{2}})\,
\delta(\rmsCoDUM    -\rmsCo_{j_{1}j_{2}})\,
\bigg]_\AVdisor.
\nonumber
\end{eqnarray}%
Here, the sum is taken over all pairs of localized particles and the various $\delta$-functions are those appropriate to their (vector or symmetric tensor) arguments.
%--
The distributed parameters, \viz,
\begin{subequations}%
\begin{eqnarray}
\MUvec_{j}&\equiv&\langle\,\Rvec_{j}\,\rangle_\AVtherm\,,
\label{eq:indmean}\\
(\rmsFl_{j})_{\cartD\cartDbar}&\equiv&\langle
(\Rvec_{j}-\MUvec_{j})_{\cartD}\,(\Rvec_{j}-\MUvec_{j})_{\cartDbar}\,\rangle_\AVtherm\,,
\label{eq:indfluc}\\
(\rmsCo_{j_{1}j_{2}})_{\cartD\cartDbar}&\equiv&\langle(\Rvec_{j_{1}}-\MUvec_{{j}_{1}})_{\cartD}\,
(\Rvec_{j_{2}}-\MUvec_{{j}_{2}})_{\cartDbar}\,\rangle_\AVtherm\,,
\label{eq:indcor}
\end{eqnarray}
\end{subequations}%
quantify the thermal motion of the localized particles in terms of
(a)~their mean positions $\MUvec_{j}\,$;
(b)~their position-fluctuation tensors $(\rmsFl_{j})_{\cartD\cartDbar}\,$; and
(c)~the fluctuation correlation tensors $(\rmsCo_{j_{1}j_{2}})_{\cartD\cartDbar}$ of pairs $\{j_{1},j_{2}\}$.
%--
Specifically, $\ProbCharsLarge$ quantifies how these random characteristics of the thermal motion of the constituents are jointly distributed, and how their distribution varies with the separation $\MUvecDUM_{1}-\MUvecDUM_{2}$ of the mean positions of pairs of particles.

Of course, how well the field correlation functions capture the heterogeneity statistics depends of the level of accuracy of the calculation of these correlation functions.
%--
Not surprisingly, if one incorporates the effects of the low-energy Goldstone fluctuations, but excludes the complementary amplitude-sector fluctuations, then the resulting approximation to the heterogeneity distribution does characterize the distribution of the position fluctuations $\rmsFlDUM$ and position-fluctuation correlations $\rmsCoDUM$, conditioned on the separation of the mean positions $\MUvecDUM_{1}-\MUvecDUM_{2}$, including how the distribution of $\rmsCoDUM$ varies with the separation.
%--
But at this level of accuracy the heterogeneity distribution does not capture any correlation between the pair of values $\rmsFlDUM_{1}$ and $\rmsFlDUM_{2}\,$, regardless of how separated $\MUvecDUM_{1}$ and $\MUvecDUM_{2}$ are.
%--
Greater accuracy in the field correlation function, provided \eg, through the incorporation of order-parameter amplitude (and not only phase) fluctuations, would enable the heterogeneity distribution to characterize the joint distribution of localization lengths for pairs of  particles, conditioned on the separation of their mean positions.

Looking beyond this example of the information encoded in the two-field correlations, correlation functions featuring three (or more) fields would supply heterogeneity distributions that characterize the thermal fluctuations and correlations amongst motions involving groups of three (or more) localized particles.
%--
Here, we have focused on the formal question of what information is contained in the field correlators.
%--
Reference~\cite{ref:UMH} describes the progress that has been made, to date, in concretely extracting the statistical information encoded in them.

%--
% \label{sec:sigma}
\mysecskip
\noindent{\bf Amorphous solidification in low dimensions\/}:
%--
There is, as we have seen, a percolation aspect to amorphous solidification, arising from the attendant formation of a system-spanning network.
%--
For percolation, the lower critical dimension -- the marginal dimension at which the transition ceases to occur -- is one.
%--
However, there is also a localization aspect to amorphous solidification, and that is associated with the spontaneous breaking of translational symmetry, suggesting a lower critical dimension of two.
%--
Indeed, as discussed in Ref.~\cite{ref:UMH}, at two dimensions, Goldstone fluctuations additively renormalize (squared) localization lengths by a term that diverges logarithmically with  the system size, even when a percolating network is present.
%--
These fluctuations restore translational symmetry and suppress the order parameter to zero, but long-ranged correlations (and presumably some form of rigidity) remain.
%--
Ongoing work with Ziqi Zhou is revealing intriguing connections in low dimensions between amorphous solidification and other settings that feature strong fluctuations, such as capillary waves of interfaces between coexisting, near-critical liquid and gas states.

%--
% \section{Broader relevance\label{sec:broader}}
\mysecskip
\noindent{\bf Could the ideas reviewed here have broader relevance?\/}:
%--
Whilst systems composed of randomly crosslinked polymers, such as rubber and chemical gels, constitute a large and technologically and conceptually important class of amorphous solids, they are by no means the only such solids to be found in nature,  synthesized in the research laboratory, or manufactured industrially.
%--
Thus, it may be useful to examine the extent to which the issues and results featuring in (and behind) the present review may be applicable to other amorphous-solid--forming systems and even to systems that are not solid but bear some resemblance to amorphous solids, at least over some range of timescales. Systems we have in mind include:
%--
(i.a)~thermal systems composed of molecules or (nonmetallic) atoms interacting \via\ nondirectional (or weakly directional) forces, which are supercooled to below their (true equilibrium) melting temperatures and cooled further still, to below their glass-forming temperatures;
%--
(i.b)~thermal systems of type~(i.a), but supercooled to temperatures between the melting temperature and the glass-forming temperature, so that they continue to flow in response to shear stress (but perhaps only exceedingly sluggishly);
%--
(ii)~continuous random network glasses comprising one or more elements (\eg, Si and O), randomly bonded to one another \via\ directional covalent bonds -- here, the statistical characteristics of the network typically play a more important role than temperature does;
%--
% [Added new (iii).]
(iii)~amorphous solids formed \via\ the random bonding (\ie, linking) of small molecules with -- importantly -- controllable densities of constraining bonds;
%--
% [Was (iii); now (iv).]
(iv)~glasses consisting of densely packed of metallic atoms (usually comprising more than one element so as to frustrate regular packing), \eg, cooled very rapidly to avoid crystallization, or irradiated with ions so as to corrupt crystalline regularity; and
%--
% [Was (iv); now (v).]
(v)~athermal systems consisting of dense packings of grains or powders. 
%--

To frame the question of the portability to such systems of results obtained in the specific setting of systems composed of randomly crosslinked polymers, let us pause to identify the features of randomly crosslinked polymer systems that enter or emerge from the theory of them that we are reviewing here. They are:
(i)~the long, flexible character of the pre-linking ingredient polymers (each of which typically hosts a large number of positional degrees of freedom);
%--
(ii)~the presence of random constraints due to crosslinking -- readily identifiable, enduring in character (and often therefore referred to as {\it chemical\/} crosslinks as opposed to {\it physical\/} crosslinks), and governed by its own statistics; and
%--
%------ POSSIBLY SHORTEN THIS SECTION / BEGIN -----------------
(iii)~the role of symmetry. Pre-crosslinking, both the system and the equilibrium state it exhibits are translationally invariant. Post-crosslinking, the laws governing the system retain this symmetry, because crosslinking constraints do not explicitly break translational invariance. However, when present in sufficient numbers, crosslinking induces spontaneous symmetry breakdown: translational invariance is lost and rigidity emerges, although the pattern of symmetry breaking is unusual, inasmuch as translational symmetry is retained macroscopically. %------ POSSIBLY SHORTEN THIS SECTION / END -----------------

%------ POSSIBLY OMIT THIS SECTION AND USE IT ELSEWHERE / BEGIN -----------------
The enduring character of the crosslinks is both conceptually important and technically empowering, as it introduces a {\it separation of timescales\/}.
%--
Conceptually, it ensures that the positional degrees of freedom (\ie, the positions of the independent segments of the polymers) have ample opportunity equilibrate within a timescale much shorter than the characteristic timescale over which the chemical bonds responsible for the crosslinking persist.
%--
Technically, this separation of scales is empowering because it opens up the possibility of treating the independent polymer segments as degrees of freedom that equilibrate (\ie, as annealing variables), and do so in the presence of a fixed collection of unvarying constraints imposed by the chemical bonds (\ie, the quenched variables).
%--
Thus, there is a window of timescales during which the network of polymers can be regarded as an equilibrated system, albeit a randomly constrained one.
%--
Said another way, there are two distinct ensembles: one explored \via\ thermal fluctuations; the other explored as one averages over the realizations of the crosslinking.
%--
The ensembles are externally defined and readily identifiable.
%--
Hence, the framework of equilibrium statistical mechanics in the presence of quenched randomness -- managed here \via\ the application of the replica technique -- is applicable and, indeed, furnishes us with a theory of the equilibrium amorphous solid state as well as the crosslinking-induced transition to this state.
%------ POSSIBLY OMIT THIS SECTION AND USE IT ELSEWHERE / END -----------------

Amongst these features, the essential ones for the theory to be unarguably applicable are: (i)~the presence of both annealing variables and quenched variables -- distinct and readily distinguishable from one another \via\ the separation of scales associated with their temporal variation; (ii)~the pattern of spontaneous symmetry-breaking, as detected and characterized \via\ the appropriate order parameter; and (iii)~the continuous nature of the transition. This continuity justifies the expansion of the Hamiltonian (in powers of the fraction of localized particles and in powers of wave-vectors) in the transition regime, where the relevant lengthscales characterizing the amorphous sold state are much larger than the intrinsic size of the entities being crosslinked. These features determine the structure of, or emerge from, the Landau-Wilson theory. (Not essential is the polymeric nature of the ingredients.) %

With these ideas in mind, we return to the question of applicability to various settings of the results reviewed here and, more generally, results obtainable from the statistical field theory of amorphous solidification.
%--
%--
\hfil\break\noindent
(i.a)~{\it Supercooled liquids below the glass-forming temperature\/}
(for reviews, see, \eg, Refs.~\cite{ref:Ediger-1996,ref:Charbonneau-2017}):
Whilst the issue of how to precisely characterize systems of this type remains an open and fascinating challenge, empirically they are amorphous solids, at least on experimentally reasonable timescales.
%--
The theory that explains the heterogeneity of the amorphous solid state in permanently constrained systems is not {\it directly\/} applicable, for the following reasons.
%--
Whilst the laws that govern the system are translationally invariant and the state it exhibits breaks that symmetry (inasmuch as detectable motions do not indicate its restoration on any experimentally accessible timescales)~\cite{Two-versions}, what differentiates glasses of this type from randomly crosslinked systems is absence of a well-defined window of timescales that is realized through the distinction between the (annealing) positional variables and the (quenched) specification of the crosslinks.
%--
Instead, the forces that restrict the motion of the constituent atoms or molecules are internally generated. Thus, it remains unclear how to identify or characterize these forces for incorporation as an element of a theoretical description; and they are likely to be connected through a continuum of timescales to the motions that do take place on experimental timescales.
%--
The absence of both (1)~a timescale window and (2)~a scheme for identifying, characterizing and incorporating the restrictions on motion whose impact seems experimentally evident are obstacles to the unarguable application to glasses of the replica theory of the amorphous solid state.
%--
Nevertheless, the similarity in the structure of glassy and crosslinked systems suggests value to examining the distribution of particle localization lengths at various temperatures below the glass transition temperature (for fit with prediction as well as for data collapse) as well as the scale-dependent elasticity (perhaps \via\ an analysis of fluctuations).
%--
It is conceivable that the various predictions made for permanently randomly constrained systems have a kind of universality that extends their validity beyond the setting that gave rise to them.
%--
And even if these actual predictions fail for glasses, the characterizing entities that they concern -- \ie, the distribution of localization lengths and any scaling of it, the emergence of rigidity with a characteristic exponent governing the shear modulus, and the scale-dependence of the shear modulus -- may turn out to be of value in settings that lie beyond their initial source.
%--
%--
\hfil\break\noindent
(i.b)~{\it Supercooled liquids above the glass-forming temperature\/}
(again, see, \eg, Refs.~\cite{ref:Ediger-1996,ref:Charbonneau-2017}):
These systems are fluid -- they develop a steady macroscopic strain rate of measurable magnitude in response to the application of a steady macroscopic stress. It seems highly likely, even absent macroscopic flow, that the constituent atoms or molecules are delocalized, wandering macroscopic distances given sufficient time.
%--
Nevertheless, for temperature histories that produce very high viscosities, one anticipates particle motions that remain well localized until very slow, collective motions \lq\lq uncage\rq\rq\ particles and permit them to roam farther.
%-
On timescales shorter than some cage renewal time, particle motions will resemble the type~(1.a) behavior discussed above, which makes it conceivable that the characteristics of the equilibrium amorphous solid state will extend to highly viscous forms of supercooled liquids above the glass-forming temperature.
%--
% Logically, a particle could be localized within a 2D plane of atoms that glides frictionally over another plane to produce a strain rate.
%--
%--
\hfil\break\noindent
(ii)~{\it Continuous random network glasses\/}:
For these systems, the connection with randomly constrained media is more direct, if one takes the covalent bonds to be permanent and the synthesis to be made from the molten state.
%--
Indeed, attempts have been made to develop a theory of network glasses by regarding them as molten atomic or molecular liquids into which permanent, directional, covalent bonds have been introduced at random, and invoking an elaboration of the replica formalism used here~\cite{ref:Shakh-1999}.
%--
Near-threshold systems, for which the link density is only modestly larger than that necessary to induce solidification (so that thermal positional fluctuations of the constituents considerably exceed the size of the unlinked atoms or molecules), should be governed by the replica approach considered here.
%--
Thus, it would be of interest to explore the localization-length distribution and elastic characteristics of such systems, and compare them with the predictions of the replica formalism.
%--
%--
% [** The next title used to read: 
%    "glass" (not amorphous solid) "formation"]
\hfil\break\noindent
(iii)~{\it Amorphous solid formation \via\ random particle-particle linking with controlled link-densities\/}:
Ozawa {\it et al\/}.~\cite{ref:OzawaEtAl-2023} have recently introduced a technique for establishing controllable densities of linking constraints (in the form of intermolecular bonds) into systems composed of small molecules.
%--
This brings the compelling opportunity to tune the resulting equilibrium amorphous solid state formed by such systems, instead of having to settle for the structural regimes of network media that kinetics delivers.
%--
In particular, it should be possible to create samples of permanently bonded small-molecule systems having crosslink densities such that the scale of particle position-fluctuations is large, and even to access the amorphous-solidification transition regime.
For such systems and in such regimes, the statistical field theory approach reviewed here should be directly relevant, as should be the semi-microscopic approach to small-molecules network media, developed in Refs.~\cite{ref:Shakh-1999}.
%--
%--
\hfil\break\noindent
% (iii)~{\it Metallic glasses\/}:
   (iv)~{\it Metallic glasses\/}:
Striking progress has recently been made in the determination of the spatial arrangement of atoms in these systems~\cite{ref:Voyles-2021,ref:Miao-2021}.
%--
However, what seems less clear is how to amplify the thermal motion of the atoms whilst maintaining the integrity of the material, so that any distribution of localization lengths would be a consequence of collective motion and -- as with supercooled liquids below the glass-forming temperature -- conceivably be governed by the predictions of the replica approach to permanently randomly constrained systems.
%--
\hfil\break\noindent
% (iv)~{\it Dense packings of grains or powders\/}:
   (v)~{\it Dense packings of grains or powders\/}
(for a review, see, \eg, Ref.~\cite{ref:Liu-Nagel-2010}):
In the macroscopically jammed regime, these are solids that are amorphous in structure: the particle locations exhibit no long-ranged spatial organization.
%--
The excitations supported in this regime include bounded motions of relatively localized, compact regions. Owing to the structural randomness, the RMS displacements of the particles will be continuously distributed (in the large-system limit).
%--
There are of course conceptual differences between jammed systems and randomly constrained thermally fluctuating systems -- preparation variations do yield an ensemble of jammed states loosely reminiscent of a percolation ensemble, but there is no analog of a thermal ensemble and, consequently, no window of timescales between alterations of quenched and annealing variables. %--
Despite the differences, it would be of interest to compare the distribution of localization lengths in the jammed state with the predictions of the theory reviewed here, not only for numerical fit but also in view of the potential for data-collapse as the packing fraction is tuned near the jamming point. Similarly, one can envision simulations in which jammed systems are subjected to shear deformations of tunable wavelength, from which the scale-dependent elastic shear modulus could be extracted.
%--
It would be of interest to determine whether the physically intuitive scenario of shear-modulus reduction due to the fraction of particles that are only weakly localized, reviewed here, also holds in the setting of jammed systems.

\mysecskip
\noindent{\bf Acknowledgments\/}:
This article was written in part at the Aspen Center for Physics, which is supported by National Science Foundation grant PHY-2210452.
%--
I am delighted to express my gratitude to the many colleagues with whom I have had the pleasure of exploring the circle of ideas reviewed here.
%--

%--
{\raggedright % begins the raggedright environment

%--
} % ends the raggedright environment
\bibliography{basename of .bib file}

\bibliography{sources}

\begin{thebibliography}{99}
\bibitem{ref:crystal}
In fact, ordering in the crystalline state is more subtle. Strictly speaking, the atoms are not localized, but the collective density pattern exhibited by them spontaneously reduces the translational symmetry, from continuous to discrete.
\bibitem{ref:knotted}
Whenever crosslinks establish closed loops of polymers (at least in three space dimensions), the inability for polymer molecules to pass through one another causes the configuration space of the molecules to decompose into disjoint volumes, each comprising a translationally invariant set of configurations that are mutually accessible.
%--
In principle, one should quench into the statistical mechanics not only the collocation constraints induced by the crosslinks, but this topological restriction according to \lq\lq knot type\rlap,\rq\rq\ too.
%-
The mathematical technology for accomplishing this is not available and, even if it were available it would be prohibitively difficult to implement.
%--
However, it seems intuitively plausible that under the coarse graining that accompanies the statistical field theory approach, the impact of knots would simply amount to an upward renormalization of the density of crosslinks.
\bibitem{Two-versions}
It is worth drawing a distinction between amorphous solids in which the particles are localized and amorphous solids in which a random particle-density profile persists and yet the individual atoms themselves may wander. Both versions would be rigid, but only in the former would particles be localized.
\end{thebibliography}

\end{document}